\documentclass[conference]{IEEEtran}
\IEEEoverridecommandlockouts
\usepackage{cite}
\usepackage{amsmath,amssymb,amsfonts}
\usepackage{algorithm}
\usepackage{algorithmic}
\usepackage{graphicx}
\usepackage{textcomp}
\usepackage{xcolor}
\usepackage{bm}
\usepackage{subcaption}
\usepackage[fleqn]{mathtools}
\usepackage{hyperref}
\def\BibTeX{{\rm B\kern-.05em{\sc i\kern-.025em b}\kern-.08em
    T\kern-.1667em\lower.7ex\hbox{E}\kern-.125emX}}

\newcommand{\loss}{\mathit{l}}

\newcommand{\expected}[2]{\underset{#2}{\mathbb{E}}[#1]}
\newcommand{\paragraphb}[1]{\vspace{0.03in} \noindent{\bf #1} }

\newcommand{\MCOMMENT}[2][.6\linewidth]{%
  \leavevmode\hfill\makebox[#1][l]{//~#2}}

\graphicspath{{./images/}}

\makeatletter
\newcommand{\linebreakand}{%
  \end{@IEEEauthorhalign}
  \hfill\mbox{}\par
  \mbox{}\hfill\begin{@IEEEauthorhalign}
}
\makeatother

\begin{document}

\title{Robust Adversarial Attacks Against \\DNN-Based Wireless Communication Systems
}

\author{\IEEEauthorblockN{Alireza Bahramali}
\IEEEauthorblockA{\textit{University of Massachusetts Amherst} \\
abahramali@cs.umass.edu}
\and
\IEEEauthorblockN{Milad Nasr}
\IEEEauthorblockA{\textit{University of Massachusetts Amherst} \\
milad@cs.umass.edu}
\and
\IEEEauthorblockN{Amir Houmansadr}
\IEEEauthorblockA{\textit{University of Massachusetts Amherst} \\
amir@cs.umass.edu}
\linebreakand
\IEEEauthorblockN{Dennis Goeckel}
\IEEEauthorblockA{\textit{University of Massachusetts Amherst} \\
dgoeckel@engine.umass.edu}
\and
\IEEEauthorblockN{Don Towsley}
\IEEEauthorblockA{\textit{University of Massachusetts Amherst} \\
towsley@cs.umass.edu}
}

\maketitle

\begin{abstract}
Deep Neural Networks (DNNs) have become prevalent
in wireless communication systems due to their promising performance.
However, similar to other DNN-based applications,
they are vulnerable to adversarial examples. In this work, we
propose an input-agnostic, undetectable, and robust adversarial
attack against DNN-based wireless communication systems in both white-box and black-box scenarios. We design tailored Universal
Adversarial Perturbations (UAPs) to perform the attack. We also
use a Generative Adversarial Network (GAN) to enforce an undetectability
constraint for our attack. Furthermore, we investigate
the robustness of our attack against countermeasures.
We show that in the presence of defense mechanisms
deployed by the communicating parties, our attack performs significantly
better compared to existing attacks against DNN-based wireless systems.
In particular, the results demonstrate that even when employing well-considered
defenses, DNN-based wireless communications are vulnerable to adversarial
attacks.
\end{abstract}


\section{Introduction}\label{intro}

Thanks to their promising classification performances, Deep Neural Networks (DNNs) have recently become a part of various wireless communication systems.
Example DNN-based wireless communication systems include
 autoencoder wireless communication~\cite{o2017introduction, nachmani2016learning, nachmani2017rnn, jiang2019turbo, liang2018iterative, gruber2017deep}, modulation recognition (radio signal classification)~\cite{o2016convolutional, west2017deep, o2018over, ramjee2019fast, meng2018automatic}, and OFDM channel estimation and signal detection~\cite{ye2017power, zhao2018deep} which leverage the high utility  of DNNs to improve the performance of their underlying wireless applications.
However, DNNs are known to be susceptible to \emph{adversarial examples},
which are small perturbations added to the inputs of a DNN causing it to misclassify the perturbed inputs.

In this paper, we study adversarial examples against wireless systems that rely on DNNs. In this setting,  the goal of an attacker is to transmit a well-crafted perturbation signal over the channel such that the underlying DNN-based wireless system (e.g., a radio signal classifier)  fails to function properly by misclassifying the perturbed signals.
Specifically, we aim at designing an  adversarial example mechanism for DNN-based wireless systems that
satisfies three key properties;
first, it should be \textbf{input-agnostic} meaning that the attacker generates the perturbation signal without any knowledge about the incoming (unknown) input signals.
This is essential as, unlike traditional targets of adversarial examples (e.g., image classification tasks),  in DNN-based wireless systems the signal to be perturbed  is not known a priori to the adversary.
Second, the perturbations should be \textbf{undetectable} in that  one should not be able to distinguish between generated adversarial perturbations and the natural noise expected from the channel.
Finally, the attack needs to be \textbf{robust} against countermeasures meaning that the defender (e.g., a wireless decoder) should not be able to remove the perturbation from the received signal.

Note  that while there exists a large body of work on adversarial examples against image classification tasks, e.g., FGSM~\cite{DBLP:journals/corr/GoodfellowSS14}, such works can not be trivially applied to the setting of wireless systems where the input signals to be perturbed are unknown to the adversary.
Some recent works~\cite{sadeghi2019physical, sadeghi2018adversarial, kim2020over, kim2020channel, restuccia2020generalized, albaseer2020performance, flowers2019communications} have investigated input-agnostic adversarial examples for  wireless systems through generating universal adversarial perturbations (UAP)~\cite{moosavi2017universal}.
However, most of the existing works~\cite{sadeghi2019physical, sadeghi2018adversarial, albaseer2020performance, kim2020over, kim2020channel} use a single-vector perturbation to attack the target wireless model which do not provide the other two key properties that we presented above, i.e., undetectability and robustness, and therefore, as we demonstrate in our paper, such adversarial examples can be easily identified and defeated (removed).
To overcome this challenge, recent works~\cite{restuccia2020generalized, flowers2019communications} use a DNN model to generate perturbations; however, they consider a white-box scenario where the adversary is aware of the target wireless DNN model.



\paragraphb{Generating input-agnostic perturbations.} In this work, we present an \emph{input-agnostic} adversarial attack against wireless communication systems that is also \emph{undetectable} and \emph{robust} to removal.
Our key approach is to model the underlying problem as an optimization problem, and solve it to obtain a
 \textit{perturbation generator model (PGM)}  that is able to generate\textemdash an extremely large number of\textemdash input-agnostic adversarial examples  vectors (i.e., UAPs) for the target wireless application.
 Therefore, instead of applying a single UAP vector (which can be easily identified and removed), in our setting the attacker picks and applies a random UAP adversarial example from a very large set of available UAPs (produced by our PGM).
 We also show that our PGM is effective in a black-box scenario where the attacker generates adversarial perturbations based on a DNN substitute model and use them to attack the original wireless DNN model.

\paragraphb{Undetectability.} We tailor the obtained PGM to specific wireless communication systems by   enforcing constraints that are specific to such systems,  therefore making the attack  undetectable and robust.
In particular,
we  use generative adversarial networks (GAN) to enforce an undetectability constraint on the UAPs generated by our PGM, and make them follow a Gaussian distribution, which is the expected noise distribution for AWGN  wireless channels. We show that by using such undetectability constraint, the PGM can completely fool a discriminator function, i.e.,  a DNN classifier that tries to distinguish between adversarial perturbations and natural Gaussian noise.
Based on our experiments, enforcing our undetectability constraint can decrease the $f1\_score$ of the discriminator from $0.99$ to $0.6$ (where an $f1\_score=0.5$ is the best undetectability as it represents random guess) with only a slight degradation in the  performance of our attack. The score can be further decreased at the cost of further performance degradation.

\paragraphb{Robustness.}
We also enforce a robustness constraint on the UAPs generated by our PGM.
Our enforced constraint aims at maximizing the distances between different UAPs generated by our PGM; this is because if the UAPs are similar, as we show, an adversary can remove their effect with the knowledge of as little as a single pilot UAP vector.
We analyze the robustness of our attack in different scenarios (Adversarial Training and perturbation subtracting) based on different amounts of knowledge available to the defender,
showing that it provides high robustness against defense techniques; by contrast, we show that a single vector UAP, as proposed in previous work~\cite{sadeghi2019physical, sadeghi2018adversarial}, can be trivially detected and removed.
Our analysis suggests that even if a  defender has knowledge about the structure of our PGM, she will not able to mitigate the effects of the attack.
As opposed to~\cite{kokalj2019adversarial} which aims at detecting adversarial perturbations, we propose defense mechanisms to actively defend against wireless adversarial attacks and remove the effect of the perturbations.

\paragraphb{Evaluation on major wireless systems. } We have implemented and evaluated our attacks on three major classes of DNN-based wireless systems, specifically, autoencoder communication systems~\cite{o2017introduction, nachmani2016learning, nachmani2017rnn, jiang2019turbo, liang2018iterative, gruber2017deep}, radio signal classification~\cite{o2016convolutional, west2017deep, o2018over, ramjee2019fast, meng2018automatic}, and OFDM channel estimation and signal detection~\cite{ye2017power, zhao2018deep}. We show that \emph{for all  three applications, our attack is highly effective in corrupting the functionality of the underlying wireless systems,  and at the same time offers strong undetectability and robustness}.
%
For instance, for the autoencoder communication system, in the presence of an adversarial training defense, our attack can increase the block-error rate (BLER) by \emph{four orders of magnitude} with a  perturbation-to-signal ratio (PSR) of $-6dB$.
However, with a similar PSR, the single vector UAP attack~\cite{sadeghi2019physical, sadeghi2018adversarial} is mainly ineffective in the presence of the same defense mechanism.
Similarly,  in the OFDM application,  our attack  results in a \emph{9X increase} in the bit error rate (while the impact of the single vector UAP is negligible).
Furthermore, our attack presents a high robustness even in the presence of a perturbation subtracting defense (as introduced), e.g., in the modulation recognition task, our attack reduces the classification accuracy from $0.69$ to $0.23$ despite the defense mechanism (by contrast, the single UAP attack is not effective as it reduces $0.69$  to only $0.67$).



\section{Background}\label{back}

\subsection{DNN-based communication Systems}
DNNs recently play an important role in wireless communication applications due to their promising performance. In this work, \emph{we focus on three major DNN-based wireless communication applications}, as introduced below:

\paragraphb{End-to-End Autoencoder Communication Systems:}
DNNs are used for end-to-end learning of communication systems using autoencoders~\cite{o2017introduction, nachmani2016learning, nachmani2017rnn}, and they can outperform the contemporary modularized design of these systems.
Such systems implement their encoders and decoders using DNNs which are able to learn the construction and reconstruction process of the information as well as the noisy environment of the physical channel.
Oshea et al.~\cite{o2017introduction} consider a communication system design as an end-to-end reconstruction task that tries to jointly optimize transmitter and receiver components in a single process. In~\cite{nachmani2017rnn} Recurrent Neural Networks (RNN) are used to decode linear block codes.

\paragraphb{Modulation Recognition:} DNNs are also used in radio signal classification or modulation recognition~\cite{o2016convolutional, west2017deep, o2018over}, which is the task of classifying the modulation type of a received radio signal to understand the type of communication scheme used in the wireless system. This can be considered as an $N$-class decision problem where the input is a complex base-band time series representation of the received signal. \cite{o2016convolutional} applies Convolutional Neural Networks (CNNs) for the complex-valued temporal radio signal domain. They use expert feature based methods instead of naively learned features to improve the performance of the classification.

\paragraphb{Signal Detection in OFDM Systems:}
\cite{ye2017power} and \cite{zhao2018deep} deploy DNNs for channel estimation and signal detection in OFDM systems in an end-to-end manner. In ~\cite{zhao2018deep}, Zhao et al. use CNNs to design an OFDM receiver outperforming the conventional OFDM receivers that are based on a Linear Minimum Mean Square Error channel estimator. Ye et al.~\cite{ye2017power} use DNNs to estimate Channel State Information (CSI) implicitly and recover the transmitted symbols directly instead of estimating CSI explicitly and detecting the transmitted symbols using the estimated CSI.

\subsection{Adversarial Examples}
An adversarial example is an adversarially crafted input that fools  a target
classifier or regression model into making incorrect classifications or predictions.
The adversary's goal is to generate adversarial examples by adding minimal perturbations to the input data attributes.
Previous works~\cite{DBLP:journals/corr/GoodfellowSS14, dong2018boosting, moosavi2017universal} have suggested several ways to generate adversarial examples. Most of the adversarial example techniques generate perturbations specific to the input, e.g., the Fast Gradient Sign Method (FGSM)~\cite{DBLP:journals/corr/GoodfellowSS14} algorithm generates adversarial perturbations based on the input and the sign of model's gradient.
However, Moosavi-Dezfooli et al.~\cite{moosavi2017universal} introduced universal adversarial perturbations (UAPs) where the adversary generates adversarial examples that are  independent of the inputs.

Similar to other DNN-based applications, DNN-based wireless applications are susceptible to adversarial attacks~\cite{hameed2019communication, flowers2019communications, delvecchio2020effects, delvecchio2020investigating, usama2019adversarial, flowers2019evaluating, sadeghi2019physical, sadeghi2018adversarial, albaseer2020performance, bair2019limitations, kim2020over, kim2020channel, shi2020generative, restuccia2020generalized}.
Flowers et al.~\cite{flowers2019evaluating} use the FGSM method to evaluate the vulnerabilities of the raw in-phase and quadrant (IQ) based automatic modulation classification task.
There is a body of work~\cite{sadeghi2019physical, sadeghi2018adversarial, kim2020over, kim2020channel, restuccia2020generalized, albaseer2020performance, flowers2019communications} concentrating on using adversarial input-agnostic technique proposed in~\cite{sadeghi2018adversarial} to attack DNN-based wireless applications, e.g.,
in~\cite{sadeghi2018adversarial}, Sadeghi and Larsson design a single UAP vector which is added to the received DNN-based modulation recognition system and causes the receiver to misclassify the modulation used by the transmitter. \cite{sadeghi2019physical} uses the same approach in an end-to-end autoencoder communication system where an attacker can craft effective physical black-box adversarial attacks to increase the error rate of the system.

As opposed to using a single vector UAP, \cite{restuccia2020generalized, flowers2019communications} use a DNN generator model to generate perturbations, e.g.,
in \cite{flowers2019communications}, Flowers et al. generate perturbations using Adversarial Transformation Networks (ARN) in a white-box scenario to evade DNN-based modulation classification systems.
Kokalj-Filipovic et al.~\cite{kokalj2019adversarial} propose two countermeasure mechanisms to detect adversarial examples in modulation classification systems based on statistical tests. One test uses Peak-to-Average-Power-Ratio (PAPR) of received signals, while another statistical test uses the Softmax outputs of the DNN classifier.

%


\section{System Model}\label{system}
We start by presenting the system models of the three DNN-based wireless applications targeted in this paper.
A general DNN-based wireless communication system consists of a transmitter, channel, and receiver.  The input of the system is a message $s \in \mathcal{M} = \{ 1,2,...,M \}$ where $M = 2^k$ is the dimension of $\mathcal{M}$ and $k$ is the number of encoded bits per message. The transmitter employs a modulation scheme and sends the modulated symbols through the channel. The receiver receives the transmitted symbols and demodulates them to reconstruct the original symbols with the minimum possible errors.
Depending on the wireless application, each part of the system behaves differently, as overviewed in the following.

\subsection{Autoencoder Communication System}

In an autoencoder communication system, the transmitter and receiver are called \textit{encoder} and \textit{decoder}, respectively, and are implemented using DNNs.
The transmitter generates a transmitted signal $x = e(s)\in \mathbb{R}^{2n}$ by applying the transformation $e:\mathcal M \rightarrow \mathbb{R}^{2N}$ to the message $s$. Note that the output of the transmitter is an $N$ dimensional complex vector which can be treated as a $2N$ dimensional real vector. Then, the generated signal $x$ is added to the channel noise, which we consider to be an additive white Gaussian noise (AWGN). Hence, the receiver receives a noisy signal $y = x + n$ and applies the transformation $d:\mathbb{R}^{2N} \rightarrow \mathcal{M}$ to create $\hat{s} = d(y)$, the reconstructed version of the message $s$.


\subsection{Modulation Recognition}
DNN-based modulation recognition can be treated as a classification problem where the input is a complex base-band time series representation of the received signal and the goal of the model is to identify the modulation type of the transmitter. Similar to  autoencoder systems, the modulated (transformed) input message $x$ is added to the channel AWGN noise $n$, and the receiver receives a noisy complex base-band signal $y = x + n$.
In this work, we will use the GNU radio ML dataset RML2016.10a~\cite{o2016radio} and its associated DNN~\cite{o2016convolutional}.
This dataset is publicly available and also enables us to compare our attack with the single vector UAP attack proposed by~\cite{sadeghi2018adversarial}.

The GNU radio ML dataset RML2016.10a contains 220000 input samples, where each sample is associated with one specific modulation scheme at a specific signal-to-noise ratio (SNR). It contains 11 different modulations: BPSK, QPSK, 8PSK, QAM16, QAM64, CPFSK, GFSK, PAM4, WBFM, AM-SSB, and AM-DSB. The samples are generated for 20 different SNR levels from -20 dB to 18 dB with a step size of 2 dB. The size of each input vector is 256, which corresponds to 128 in-phase and 128 quadrature components. Half of the samples are considered as the training set and the other half as the test set.

\subsection{OFDM Channel Estimation and Signal Detection}
In an OFDM system, at the transmitter side the transmitted symbols and pilot signals are converted into a parallel data stream. Then, the inverse discrete cosine transform (IDFT) converts the data stream from the frequency domain to the time domain with a cyclic prefix (CP) inserted to mitigate the inter-symbol interference (ISI). The length of the CP should be no shorter than the maximum delay spread of the channel.
Based on Ye et al.~\cite{ye2017power}, we consider a  sample-spaced multi-path channel described by the complex random vector $h$. On the receiver side, the received signal can be expressed as $y = x \circledast h + n$, where $\circledast$ denotes circular convolution while $x$ and $n$ represent the transmitted signal and the AWGN noise of the channel, respectively.
At the receiver of the OFDM system, the frequency domain received signal is obtained after removing the CP and performing a DFT.

We assume that the DNN model takes as input the received data consisting of one pilot block and one data block, and reconstructs the transmitted data in an end-to-end manner. To be consistent with~\cite{ye2017power}, we consider  64 sub-carriers and a CP of length 16. Also, we use 64 pilots in each frame for channel estimation. Hence, the size of the input vector is 256, where the first 128 samples are the in-phase and quadrant components of the pilot block, and the second 128 samples are the in-phase and quadrant components of the following data block.
We use \cite{ye2017power}'s fully connected DNN model.


\section{Attack Model}\label{attack-model}
In all of the mentioned wireless applications, the goal of the \textbf{attacker}
\footnote{We use ``attacker'' and ``adversary'' interchangeably.}
is to transmit a well-designed perturbation signal over the channel such that the underlying DNN-based model misclassifies.
The attacker acts as a jammer and transmits a well-designed input-agnostic perturbation over the channel. The generated perturbations are added to the transmitted signals and AWGN over-the-air. The receiver receives the perturbed signal and applies the target DNN-based model to it.
Note that in the OFDM system, the attacker adds the generated perturbation to each frame containing a pilot block and a data block.
We consider a strong attacker in white-box and black-box scenarios.
In white-box scenario, the attacker is aware of the underlying DNN-based model, while in black-box setting the adversary has no or limited knowledge of the underlying DNN-based model.
Due to the challenges for the attacker to obtain robust phase synchronization with the transmitter at the receiver, which would likely require tight coordination with the communicating nodes, we assume that the perturbation generated by the attacker is subject to a random phase shift on the channel relative to the transmitter's signal.

\begin{figure}[t]
  \centering
  \includegraphics[width = 0.8\linewidth]{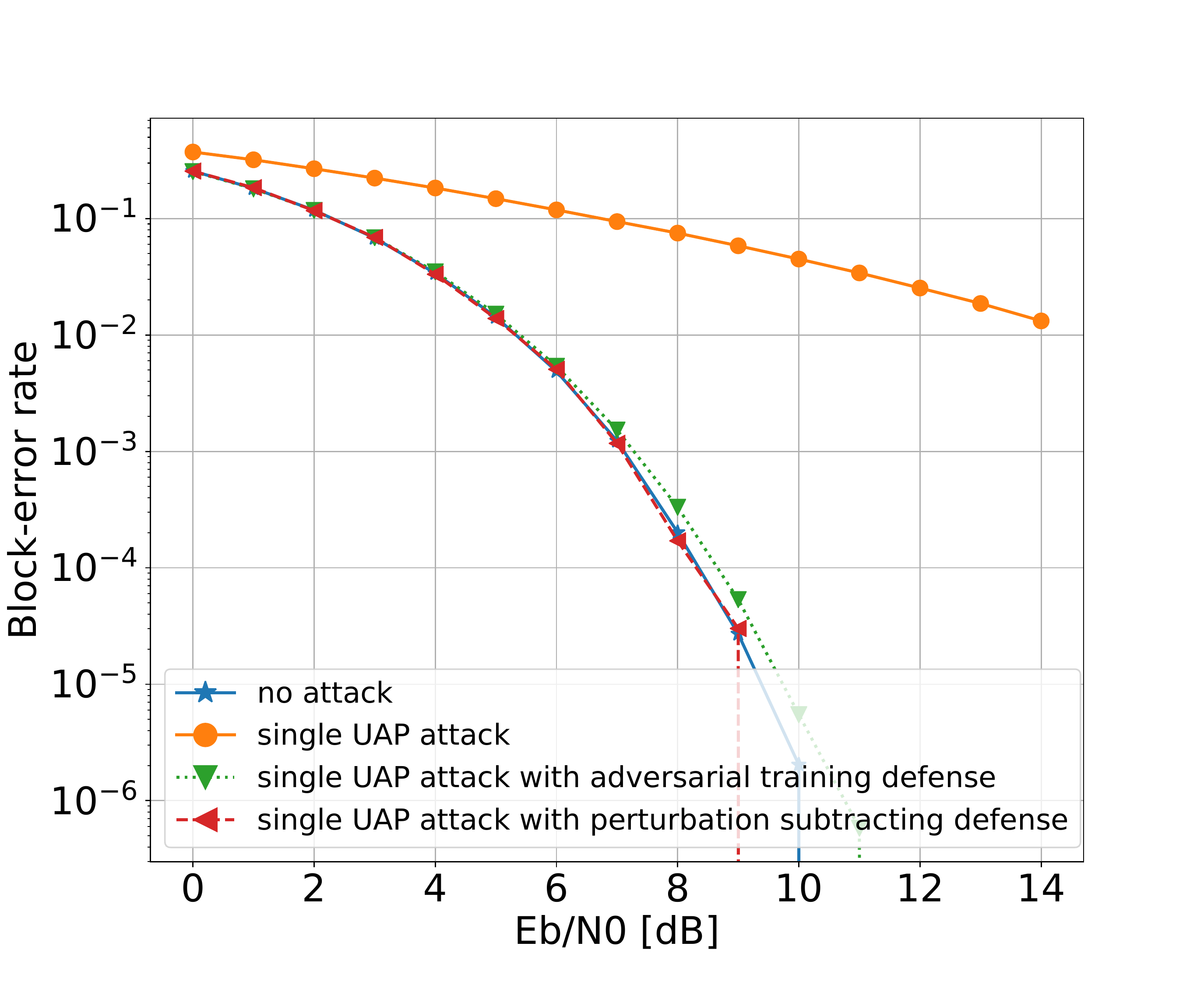}
  \caption{The single vector UAP attack~\cite{sadeghi2019physical} can be easily defeated using our two defense mechanisms.
}
  \label{fig:defense-single}
\end{figure}

As mentioned in Section~\ref{back}, in the underlying wireless applications,  the perturbation signal needs to be transmitted over-the-air, and therefore the perturbation signal should be input-agnostic i.e., universal (UAP).
This allows the attacker to generate perturbation signals with no need to knowing the upcoming wireless signals.
While
some recent works~\cite{sadeghi2019physical, sadeghi2018adversarial}  has investigated such UAPs, they are easily detectable as the attacker uses a single perturbation vector. Such perturbation vector can be inferred by the defender (e.g., through  pilot signals) and consequently subtracted from the jammed signals. We demonstrate this through two defense mechanisms (namely,  adversarial training and noise subtracting defenses as presented in Section~\ref{defense}).
As shown in Figure~\ref{fig:defense-single}, both of our defense mechanisms can easily defeat a single vector UAP as proposed by~\cite{sadeghi2019physical, sadeghi2018adversarial}.

Therefore, instead of designing a single UAP, our attacker learns the parameters of a PGM that generates separate perturbation vectors without any knowledge of the input.
Using a PGM instead of a single noise vector provides the attacker with a large set of perturbations, and we can use existing optimization techniques such as Adam~\cite{adam} to find the perturbations.


\section{Our Perturbation Generator Model}\label{model}
\begin{figure*}[t]
  \centering
  \includegraphics[width = 0.75\linewidth]{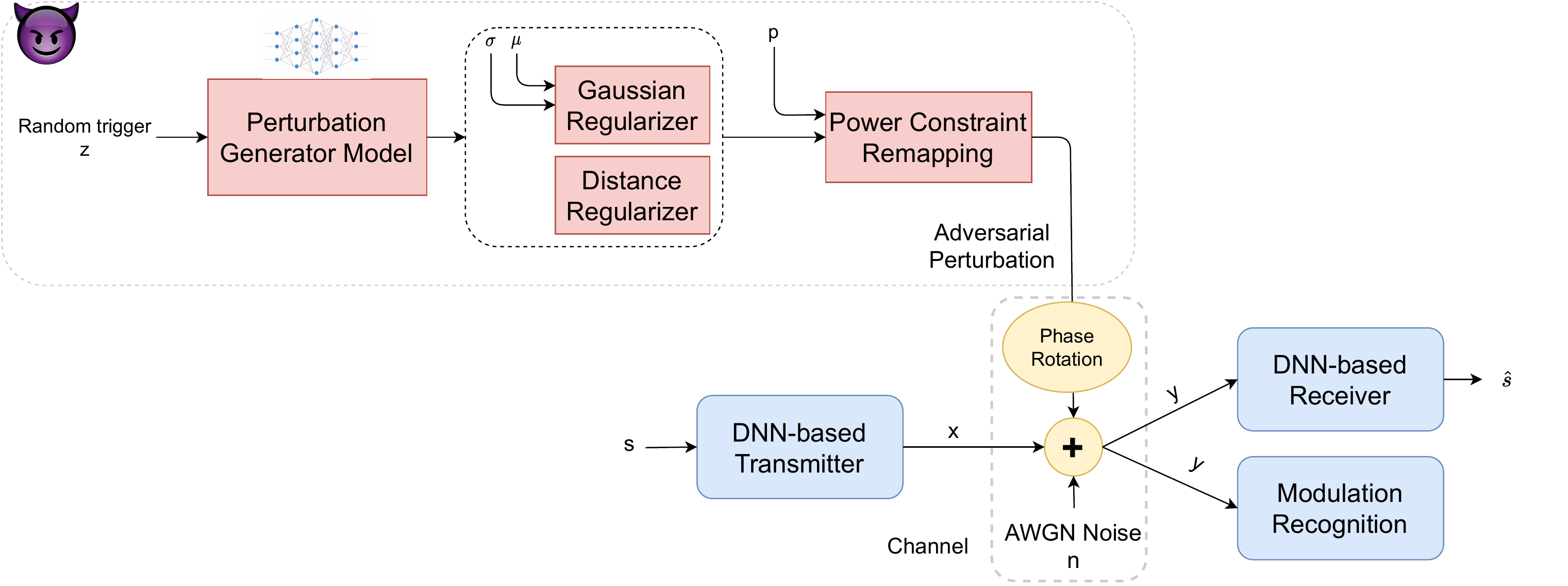}
  \caption{Our attack setting}
  \label{fig:setting}
\end{figure*}

In this section, we provide details  on how our attack is performed using a PGM. Figure~\ref{fig:setting} illustrates the process.

\subsection{General Formulation}
We formulate the universal adversarial perturbation problem in a wireless communication system as the following problem:
\begin{align}\label{eq:adv_main}
   \nonumber&\arg \min_{\delta} \; ||\delta||_{2} \\ &\text{s.t.} \; \;  \forall y \in D : f(y+\delta) \neq f(y)
\end{align}
where $y$ is the transmitted signal added by the AWGN noise ($y=x+n$) and $f$ is the underlying DNN-based function in the wireless system.
The objective is to find a minimal (perturbation with minimum power) universal perturbation vector, $\delta$, such that when added to an \emph{arbitrary} input from a target input domain $D$, it will cause the underlying DNN-based model $f(.)$ to misclassify.
Note that one cannot find a closed-form solution for this optimization problem since the DNN-based model $f(.)$ is a non-convex machine learning model, i.e., a deep neural network.
Therefore, \eqref{eq:adv_main} can be formulated as follows to numerically solve the problem using empirical approximation techniques:
\begin{align}\label{eq:erm_untargted}
    &\arg \max_{\delta}   \sum_{y \in \mathcal{D}}  \loss(f(y+\delta),f(y))
\end{align}
where $\loss$ is the DNN-based model loss function and $\mathcal{D}\subset D$ is the  attacker's network training dataset.

As mentioned above, instead of learning a single UAP as suggested by~\cite{sadeghi2019physical, sadeghi2018adversarial}, we aim at learning the parameters of a PGM $G$ to be able to generate UAPs without any knowledge of the system input. This generator model $G$ will generate UAP vectors when provided with a random \emph{trigger} parameter $z$ (we denote the corresponding adversarial perturbation as $\delta_z=G(z)$), i.e., we can generate different perturbations on different random $z$'s.
Therefore, the goal of our optimization problem is
  to optimize the parameters of the PGM $G$  (as opposed to optimizing a UAP $\delta$  in \cite{sadeghi2019physical}).
  Hence, we formulate our optimization problem as:
\begin{align}\label{eq:erm_untargted}
    & \arg \max_{G}   \expected{ \sum_{y \in \mathcal{D}}  \loss(f(y+G(z)),f(y))}{ z \sim uniform (0,1)}
\end{align}

 We can use existing optimization techniques (e.g., Adam~\cite{adam}) to  solve this problem.
 In each iteration of  the training, our algorithm selects a batch from the training dataset and a random trigger $z$,  then computes the objective function.

 Algorithm~\ref{alg:main} summarizes our approach to generate UAPs. In each iteration, Algorithm~\ref{alg:main} computes the gradient of the objective function w.r.t. the perturbation for given inputs, and optimizes it by moving in the direction of the gradient. The algorithm enforces the underlying  constraints of the wireless system using various remapping and regularization functions which will be discussed in the following sections. We use the iterative mini-batch stochastic gradient ascent~\cite{goodfellow2016deep} technique.

 \begin{algorithm}[t!]
     \caption{Generating UAPs using PGM}
     \label{alg:main}
    \begin{algorithmic}
      \STATE $\mathcal{D} \gets$ adversary training data
      \STATE $f \gets$ DNN-based model
      \STATE $y \gets$ training input
      \STATE $\mathcal{L}_{f} \gets $ DNN-based loss function
      \STATE $\mathcal{M} \gets$ domain remapping function
      \STATE $\mathcal{R} \gets$ domain regularizations function
      \STATE $G(z) \gets$ initialize the blind adversarial perturbation model parameters ($\theta_G$)
      \STATE $T \gets $ epochs
      \FOR{epoch $t \in \{ 1 \cdots T \}$  }
          \FORALL{mini-batch $b_i$ in $\mathcal{D}$}
              \STATE $z \sim$ Uniform
              \STATE Rotate $\mathcal{M}(y,G(z))$ based on the channel phase shift
              \STATE $J = -(\frac{1}{|b_i|} \sum_{\bm{x}  \in b_i, } \loss (f(\mathcal{M}(y,G(z))),f(x)) ) + \mathcal{R}(G(z))$
              \STATE  Update $G$ to minimize $J$
          \ENDFOR
      \ENDFOR
      \RETURN $G$
     \end{algorithmic}
 \end{algorithm}

\subsection{Incorporating Power Constraint}
As our first constraint on UAPs, we enforce an upper-bound  on the attacker's perturbation power.
This constraint defines the maximum power the generated UAPs can reach. To enforce this power constraint, we use a \textit{remapping function} $\mathcal{M}$ while creating the UAP. A remapping function adjusts the perturbed signal so it complies with the power constraint. Therefore, we reformulate our optimization problem by including the remapping function $\mathcal{M}$:

\begin{align}\label{eq:erm_untargted}
     & \arg \max_{G}   \expected{ \sum_{y \in \mathcal{D}}  \loss(f(\mathcal{M}(y,G(z))),f(y))}{ z \sim uniform (0,1)}
\end{align}

Each time we want to create the perturbation signal, we check its power and if it violates the constraint, we normalize it to meet the power constraint. The following shows the remapping function we used to preserve the power constraint of the perturbations in the target wireless applications.

\[ \mathcal{M}(y, G(z), p) = y +
 \begin{cases}
    \sqrt{p}\frac{G(z)}{||G(z)||_{2}},  \; \; &||G(z)||^{2}_{2} > p,  \\
   G(z), \; \;  &||G(z)||^{2}_{2} \leq p.
\end{cases} \]

\subsection{Incorporating Undetectability Constraint}
As mentioned earlier, the adversarial perturbation is not noise but a deliberately optimized vector in the feature space of the input domain, and hence is easily distinguishable from the expected behavior of noise in the communication system environment. To make our adversarial perturbation undetectable, we enforce a statistical behavior (such as Gaussian behavior) which is expected from the physical channel of a communication system on our adversarial perturbations. We use a \textit{regularizer} $\mathcal{R}$ in the training process of our PGM to enforce a Gaussian distribution for the perturbation. To do this, we use a generative adversarial network (GAN)~\cite{NIPS2014}: we design a discriminator model $D(G(z))$ which tries to distinguish the generated perturbations from a Gaussian distribution. Then we use this discriminator as our regularizer function to make the distribution of the crafted perturbations similar to Gaussian distribution. We simultaneously train the blind perturbation model and the discriminator model. Hence, we rewrite~\ref{eq:erm_untargted} as follows:

\begin{align}\label{eq:erm_untargted_gan}
   \nonumber\arg \max_{G} \expected{ ( \sum_{y \in \mathcal{D}} \loss (f(\mathcal{M}(y,G(z))),f(y)) ) \\ + \alpha\mathcal{R}(G(z))}{z \sim uniform (0,1)}
\end{align}
where $\alpha$ shows the weight of the regularizer in comparison to the main objective function.
Algorithm~\ref{alg:normal} shows details of our technique to generate Gaussian behavior enforced perturbations with an average $\mu$ and standard deviation $\sigma$. Note that we use Gaussian distribution as the desired distribution for our noise since it is expected from the environment where a wireless communication system operates and it cannot be distinguished from a normal AWGN noise of the channel.

\begin{algorithm}[t!]
    \caption{GAN-based noise regularizer}
    \label{alg:normal}
   \begin{algorithmic}
       \STATE $\mathcal{D} \gets$ training data
       \STATE $f \gets$ DNN-based model
       \STATE $G \gets$ PGM
       \STATE $D \gets$ discriminator model
       \STATE $\mu,\sigma^2 \gets$ target desired Gaussian distribution parameters
       \FOR{$t \in \{1,2,\cdots,T\}$}
	   \STATE $z' \sim \text{Gaussian}(\mu,\sigma^2)$
	   \STATE $z \sim \text{Uniform}$
	   \STATE train $D$ on $G(z)$ with label 1 and $z'$ with label 0
	   \STATE train $G$ on $\mathcal{D}^{S}$ using regularizer $\mathcal{R}$
       \ENDFOR
       \RETURN $G$
    \end{algorithmic}
\end{algorithm}

\subsection{Incorporating Robustness Constraint}
As mentioned earlier, using a PGM instead of a single UAP to perform the adversarial attack provides the adversary an extremely large set of perturbations. This makes the attack more robust against countermeasures compared to the single vector UAP attack. However, if the generated perturbations are very similar to each other, an ad-hoc defender (will be discussed in Section~\ref{defense}) can use pilot signals to obtain a close estimate of the perturbations.
To prevent this, we can force Algorithm~\ref{alg:main} to generate non-similar perturbations. To this aim, we add the $l_2$ distance between consecutively generated perturbations as a regularizer to the objective function. Hence, in the training process, our model tries to maximize the distance between perturbations.
In Section~\ref{defense}, we see that incorporating this constraint can prevent an ad-hoc defender from making a close estimation of the generated perturbations.

\subsection{Incorporating Channel Phase Rotation}
As mentioned in Section~\ref{attack-model}, to model the lack of phase synchronization between the attacker and the transmitter, we add a relative random phase to the perturbation generated by the perturbation model and rotate its signal.
For each adversarial perturbation, we generate a random phase $\theta$ and rotate the perturbation based on it.
Note that the channel effect is applied on the perturbation after applying all of the mentioned constraint on the perturbation.
Assume $p = \mathcal{M}(x, G(z))$ is a perturbation generated by the PGM after applying the power constraint, and $p_R$ and $p_I$ are the real and imaginary parts of it respectively. Using the random phase shift caused by the channel the rotated perturbation can be derived as follows:
\begin{align}
  &\nonumber \text{For all}\; i=1,2,\cdots,N\text{:} \\
  &\begin{cases}
    p_{i,R}^\prime = p_{i,R}\cos(\theta) - p_{i,I}\sin(\theta) \\
    p_{i,I}^\prime = p_{i,I}\cos(\theta) + p_{i,R}\sin(\theta)
  \end{cases}
\end{align}

Where $N$ is the length of the perturbation, $p_{R}^\prime$ and $p_{I}^\prime$ are the real and imaginary parts of the rotated perturbation.

%
%
%


\section{Experimental Setup}

For the three target  wireless applications, we use the same setup as their original papers~\cite{o2017introduction, o2016convolutional, ye2017power}. Table~\ref{tab:params} summarizes the settings of our experiment for each application.
The target models for each application are the same as the ones proposed in their original papers. Table~\ref{tab:models} illustrates the parameters of each mode.
To enable a benchmark for comparison, we obtain the code of the UAP adversarial attack proposed by~\cite{sadeghi2019physical, sadeghi2018adversarial} and transform it from Tensorflow to Pytorch.
 We then compare the results of our adversarial attack using a PGM with the UAP adversarial attack. However, appear to be the first to apply the adversarial attack on the OFDM channel estimation and signal detection task; hence, there is no baseline to compare our attack for the OFDM system.

We use fully connected layers for the PGM with different numbers of hidden layers and different numbers of neurons in each layer based on the wireless application. Table~\ref{tab:params} contains the details of the structure used for each PGM.

For the discriminator model mentioned in Section~\ref{model}, we use a fully connected DNN model with one hidden layer of size 50 and a ReLU activation function. The discriminator generates a single output which can be interpreted as the probability of being a Gaussian distribution for a signal. In the training process of our discriminator, we set $\mu$ and $\sigma$ to the average of the mean and standard deviation of the generated perturbations. To train the discriminator we use the Adam~\cite{adam} optimizer with a learning rate of $10^{-5}$.

\begin{table*}[!t]
\centering
\caption{Details of the perturbation generation model in each wireless application}
\begin{tabular}{|c|c|c|c|}
  \cline{2-4}
  \multicolumn{1}{c|}{}& \textbf{Autoencoder End-to-End Communication} & \textbf{Modulation Recognition} & \textbf{OFDM Channel Estimation} \\
  \hline
  input size & $2\times 7$ & $2 \times 128$ & 256\\
  \hline
  hidden layers sizes & 100 & 5000, 1000 & 5000, 1000\\
  \hline
  hidden layers activations & ReLU & Leaky ReLU, Leaky ReLU & Leaky ReLU, Leaky ReLU\\
  \hline
  loss function & Cross Entropy & Cross Entropy & MSE \\
  \hline
  metric & Block-Error Rate (BLER) & Accuracy & Bit Error Rate (BER)\\
  \hline
  optimizer & Adam & Adam & Adam\\
  \hline
  learning rate & $10^{-4}$ & $10^{-3}$ & $10^{-2}$\\
  \hline
\end{tabular}
\label{tab:params}
\end{table*}

\begin{table*}[!t]
\centering
\caption{Details of the target models for each wireless application}
\begin{tabular}{|c|cc|c|c|}
  \cline{2-5}
  \multicolumn{1}{c|}{}& \textbf{Autoencoder Encoder} & \textbf{Autoencoder Decoder} & \textbf{Modulation Recognition} & \textbf{OFDM Channel Estimation} \\
  \hline
  input size & 16 & $ 2 \times 7$ & $1\times 1\times 256$ & 256\\
  \hline
  hidden layers sizes & 16 & 16 & 10560, 256 & 500, 250, 120\\
  \hline
  hidden layers activations & eLU & ReLU & Leaky ReLU, Softmax & ReLU, ReLU, ReLU \\
  \hline
  kernels & \_ & \_ & 256, 80 & \_ \\
  \hline
  kernel size & \_ & \_ & $(1,3)$, $(2,3)$ & \_ \\
  \hline
  stride & \_ & \_ & $(1,1)$, $(1,1)$ & \_ \\
  \hline
  padding & \_ & \_ & $(0,2)$, $(0,2)$ & \_ \\
  \hline
  convolutional activations & \_ & \_ & Leaky ReLU, Leaky ReLU & \_ \\
  \hline
  output size & $2\times 7$ & 16 & 11 & 16 \\
  \hline
\end{tabular}
\label{tab:models}
\end{table*}




\section{Experimental Results}\label{exp}
In this section, first we evaluate our attack against three target wireless applications without any undetectability constraint. We also compare our attack with the single vector UAP.
As mentioned in Section~\ref{attack-model}, to consider the channel effect, we add a relative random phase shift to our perturbations.
Second, we evaluate our attack while enforcing the undetectability constraint mentioned in Section~\ref{model} on the autoencoder communication system.
Figure~\ref{fig:attack} shows the performance of our attack and the single vector UAP attack for three target applications. In the following we analyze the results for each application.

\subsection{Performance Without the Undetectabilty Constraint}
\begin{figure*}
  \begin{subfigure}{0.33\textwidth}
    \centering
    \includegraphics[width=\linewidth, trim={1cm, 0cm, 1cm, 1cm}]{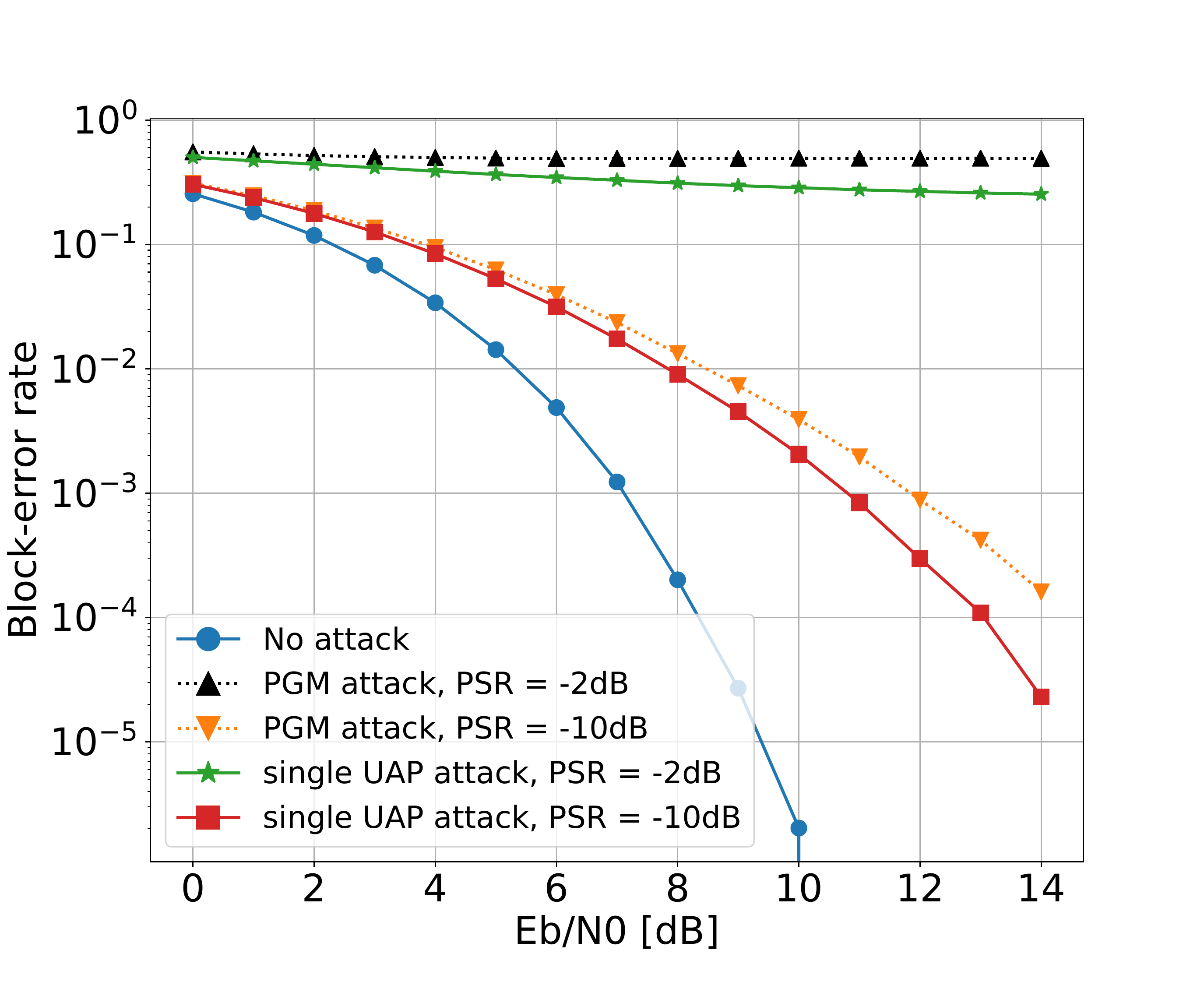}
    \caption{\footnotesize Autoencoder communication system}
    \label{fig:attack-autoencoder}
  \end{subfigure}
  \begin{subfigure}{0.33\textwidth}
    \centering
    \includegraphics[width=\linewidth, trim={1cm, 0cm, 1cm, 1cm}]{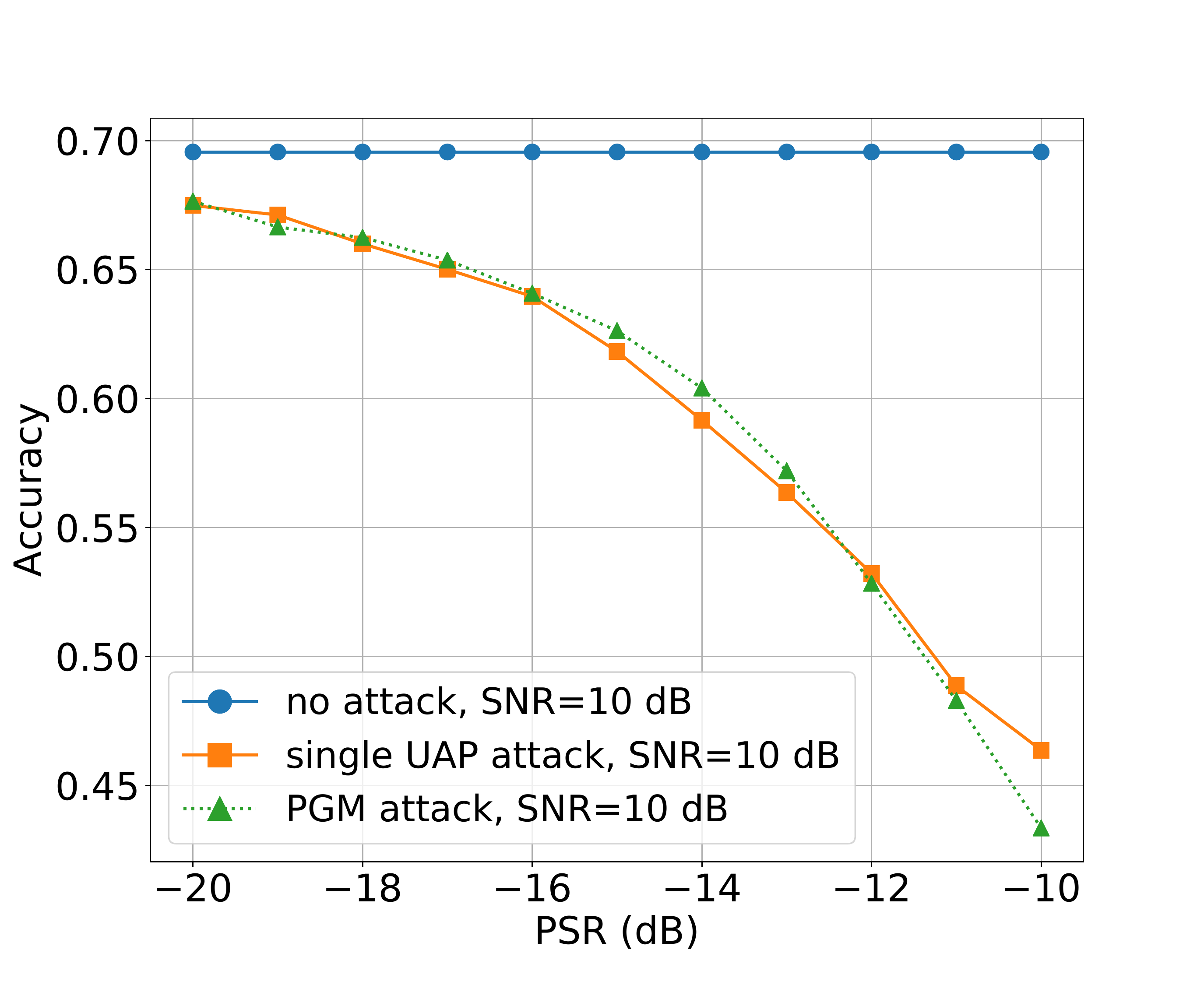}
    \caption{\footnotesize Modulation recognition system}
    \label{fig:attack-modulation}
  \end{subfigure}
  \begin{subfigure}{0.33\textwidth}
    \centering
    \includegraphics[width=\linewidth, trim={1cm, 0cm, 1cm, 1cm}]{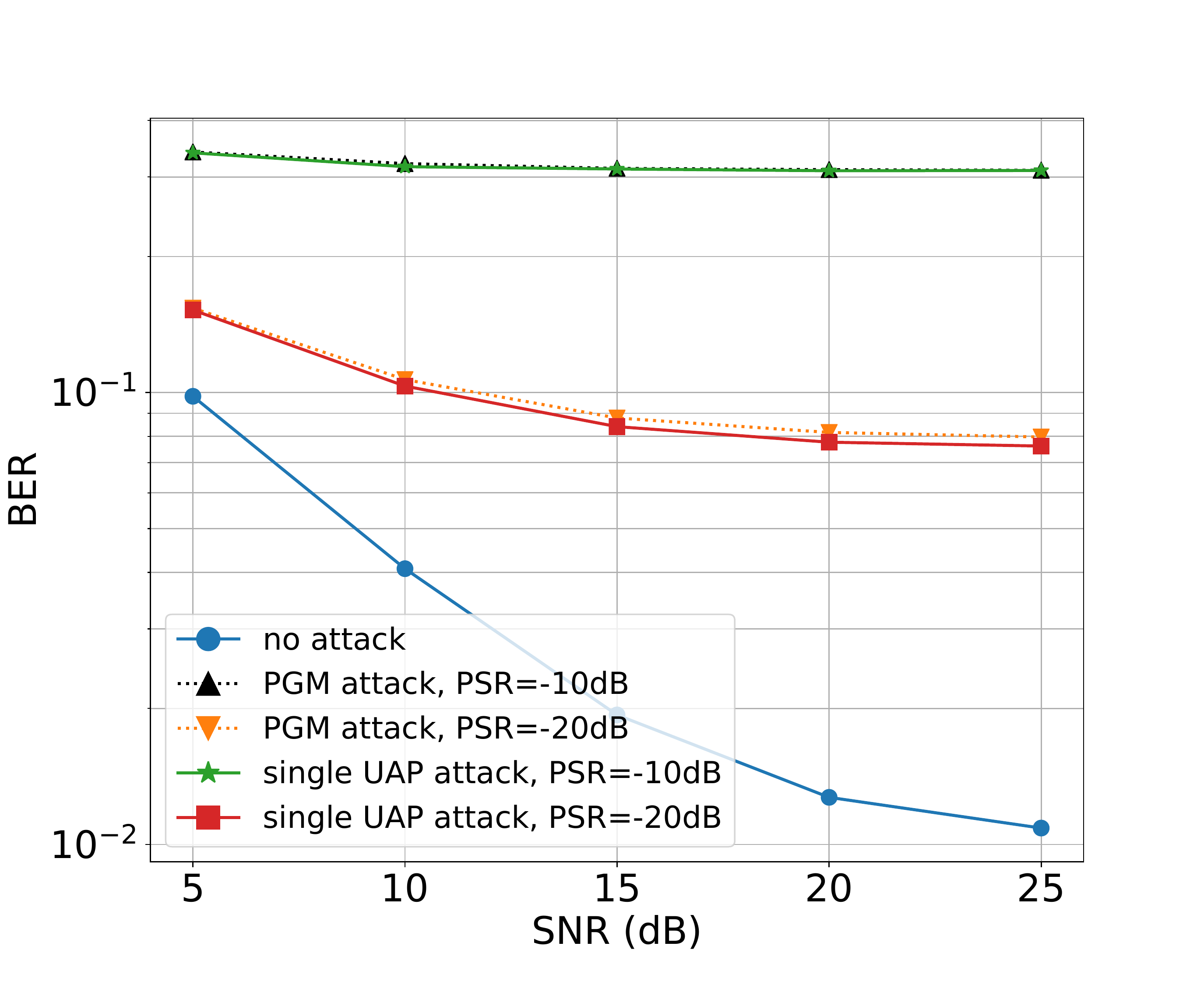}
    \caption{\footnotesize OFDM system}
    \label{fig:attack-ofdm}
  \end{subfigure}
  \caption{Performance of our attack and the single vector UAP attack for the three target wireless systems.}
  \label{fig:attack}
\end{figure*}

\paragraphb{Autoencoder Communication System:}
Figure~\ref{fig:attack-autoencoder} shows the BLER performance of the autoencoder communication system under adversarial attack while using a PGM and a single vector UAP attack proposed in~\cite{sadeghi2019physical}.
To be consistent with~\cite{o2017introduction}, we set $N=7$ and $k=4$.
We sweep $Eb/N0$ from $0dB$ to $14dB$ with steps of $1dB$, and for each value, we calculate the BLER of the autoencoder system.
Similar to~\cite{sadeghi2019physical}, to compare the power of the adversarial perturbation at the receiver with the received signal, we introduce a parameter, the perturbation-to-signal ratio (PSR), which equals the ratio of the received perturbation power to the received signal power.

We see that for different PSR ratios, using a PGM to generate the UAPs increases the performance of the adversarial attack in comparison to learning a single UAP. As mentioned in Section~\ref{model}, the reason is that by learning the parameters of a generator model instead of learning a single perturbation vector, we can leverage existing learning techniques (such as momentum-based ones) to ease the process of learning and prevent common learning problems such as getting stuck in local minima.


\paragraphb{Modulation Recognition:}
Figure~\ref{fig:attack-modulation} shows the performance of our adversarial attack using a PGM against the modulation recognition application over different values of PSR. The GNU~\cite{o2016radio} dataset contains samples for different values of SNR; however, we apply our attack using samples with SNR of 10 dB. Figure~\ref{fig:attack-modulation} also shows the comparison between our attack and the single vector UAP attack proposed by~\cite{sadeghi2018adversarial} against modulation recognition. We see that using a PGM does not have an advantage over a single vector UAP method in terms of attack performance if the wireless system does not employ defenses. In Section~\ref{defense}, we will demonstrate that the PGM is significantly more robust against defense mechanisms compared to the single vector UAP technique.


\paragraphb{OFDM Channel Estimation and Signal Detection: }
Similar to the two above applications, we apply each of the single vector UAP attack and adversarial PGM on the DNN-based OFDM system proposed by~\cite{ye2017power}. Because there is no reported prior adversarial attack on the DNN-based OFDM system, we do not have any baseline for comparison. Figure~\ref{fig:attack-ofdm} shows the Bit Error Rate (BER) of the OFDM system against the two mentioned adversarial attacks. The SNR is varied from $5dB$ to $25dB$ and we evaluate our attack for two values of PSR, $-10dB$ and $-20dB$. We see that using a PGM improves the adversarial attack slightly compared to the single vector UAP attack if the wireless system does not employ any defenses. However, in Section~\ref{defense}, we will see that using our PGM makes the attack significantly more robust against possible defenses.


\subsection{Performance With the Undetectability Constraint}
In this section, we evaluate our attack while enforcing an undetectability constraint using a GAN in the autoencoder communication system. Note that this technique is easily applicable to other systems since it uses a discriminator network independent of the underlying DNN model in the wireless application.
To investigate the undetectability of the generated perturbations, we train our discriminator in two scenarios: first, we train our discriminator to be able to distinguish between adversarial noise and a natural Gaussian noise without enforcing an undetectability constraint on the adversarial generator model. In the second scenario, we train the discriminator while enforcing a Gaussian distribution on our adversarial noise in the PGM. The evaluation metric for the discriminator is the $f1\_score$ which can be interpreted as a weighted average of the precision and recall metrics.

\begin{figure}[t]
  \centering
  \includegraphics[width = 0.8\linewidth]{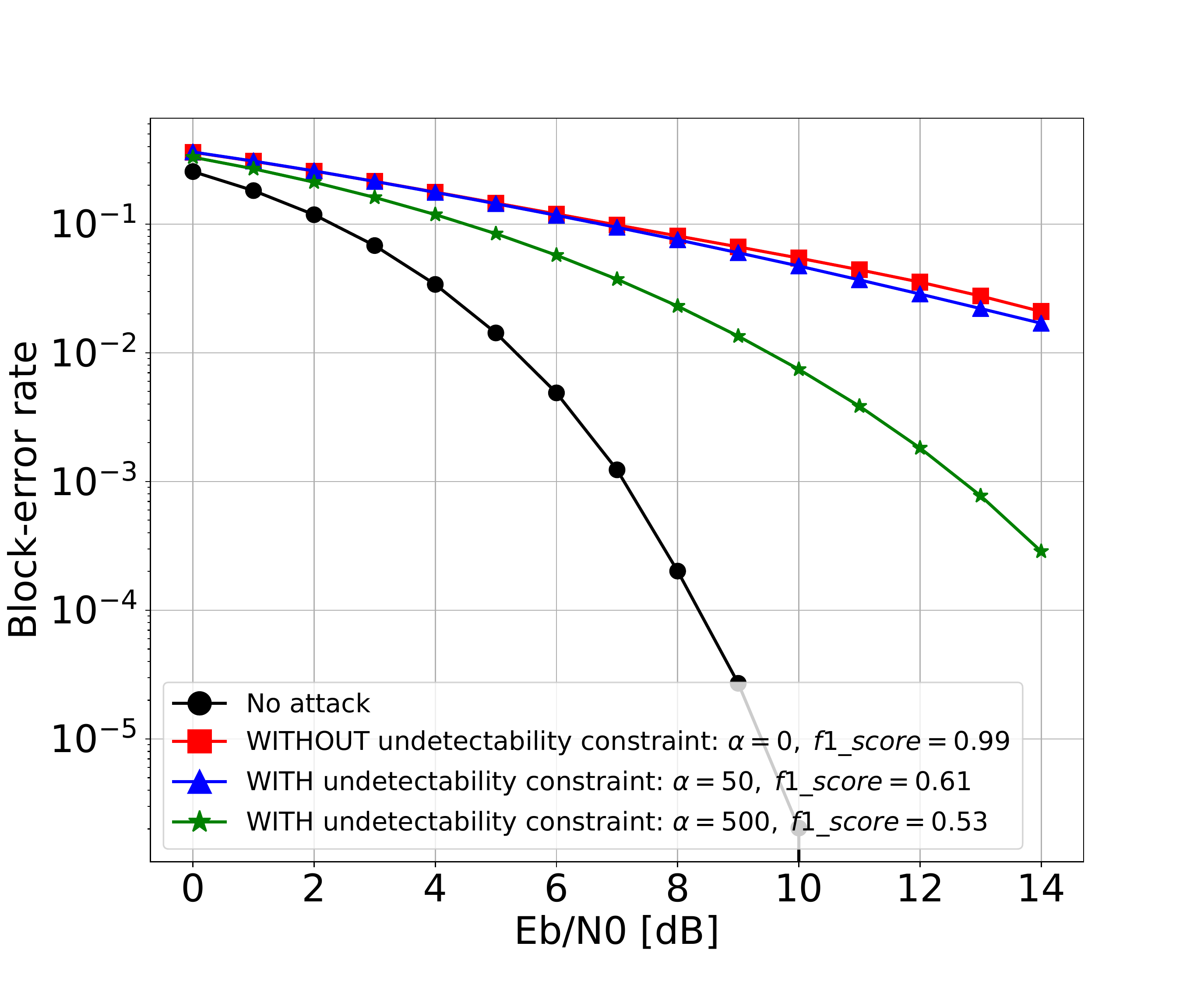}
  \caption{Performance of the autoencoder communication system against PGM attack with and without the undetectability constraint and the corresponding $f1\_score$ of the discriminator.}
  \label{fig:gan}
\end{figure}

Figure~\ref{fig:gan} shows the performance of the discriminator as well as the generator in these two scenarios for different values of $\alpha$ (denoting the strength of the undetectability constraint used in Equation~\ref{eq:erm_untargted_gan}) while the PSR of the generated perturbations is $-6dB$.
Note that since the $f1\_score$ is almost the same for different values of $Eb/N0$, we consider the average $f1\_score$ over different values of $Eb/N0$.
We see that without enforcing the undetectability constraint in the PGM ($\alpha=0$), the discriminator is able to distinguish between generated adversarial noise and Gaussian noise: the $f1\_score$ is nearly 1. On the other hand, when we enforce an undetectability constraint in the PGM with $\alpha=50$, the average $f1\_score$ is $0.61$, which means that the discriminator misclassifies the generated perturbations as Gaussian noise to some extent while the performance of our attack decreases slightly compared to a scenario where there is no undetectability constraint. By increasing $\alpha$ to 500, the average $f1\_score$ becomes $0.53$ making our generated noise more undetectable since the discriminator cannot distinguish between Gaussian noise and adversarial noise. On the other hand with $\alpha=500$, our attack performs much worse than the case with $\alpha=50$; however, it still degrades the performance of the autoencoder significantly.
\textbf{Therefore, we show that by enforcing an undetectability constraint, we can achieve high undetectability with only a small degradation in the performance of the attack.}


\subsection{Performance With the Black-box Setting}
As mentioned in Section~\ref{attack-model}, in the white-box setting we assume that the attacker is aware of the underlying DNN-based model including its structure and parameters.
In this part, we evaluate our PGM in the black-box setting where the attacker does not have any knowledge about the underlying DNN model.
Instead, the attacker uses its own substitute model and then design a white-box attack for it, as it has the perfect knowledge of the substitute model.
The attacker then uses the crafted perturbations to attack the original unknown underlying DNN model. This is called a black-box adversarial attack~\cite{papernot2016transferability}.

\begin{figure*}
  \begin{subfigure}{0.33\textwidth}
    \centering
    \includegraphics[width=\linewidth, trim={1cm, 0cm, 1cm, 1cm}]{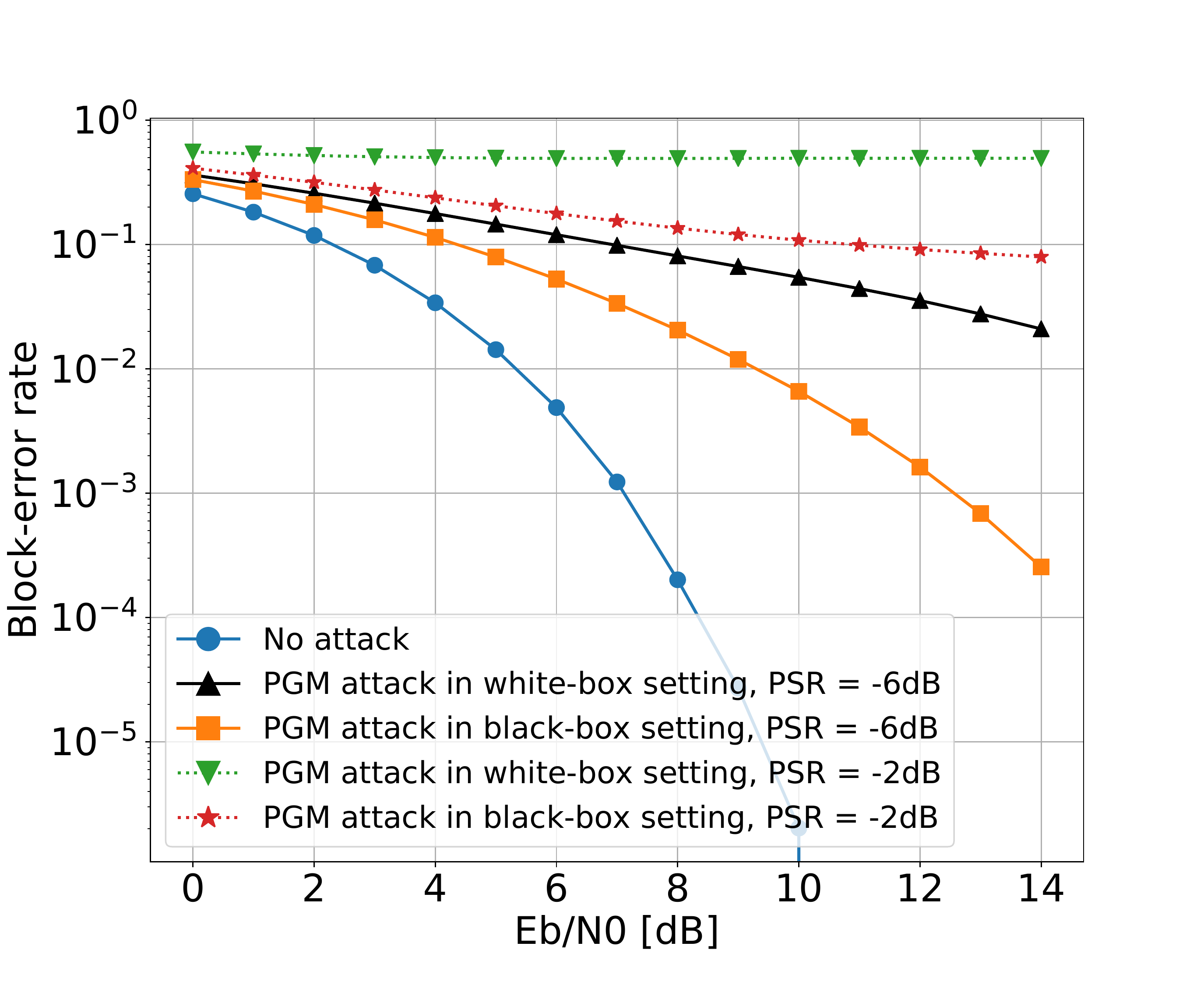}
    \caption{\footnotesize Autoencoder communication system}
    \label{fig:attack-autoencoder-bb}
  \end{subfigure}
  \begin{subfigure}{0.33\textwidth}
    \centering
    \includegraphics[width=\linewidth, trim={1cm, 0cm, 1cm, 1cm}]{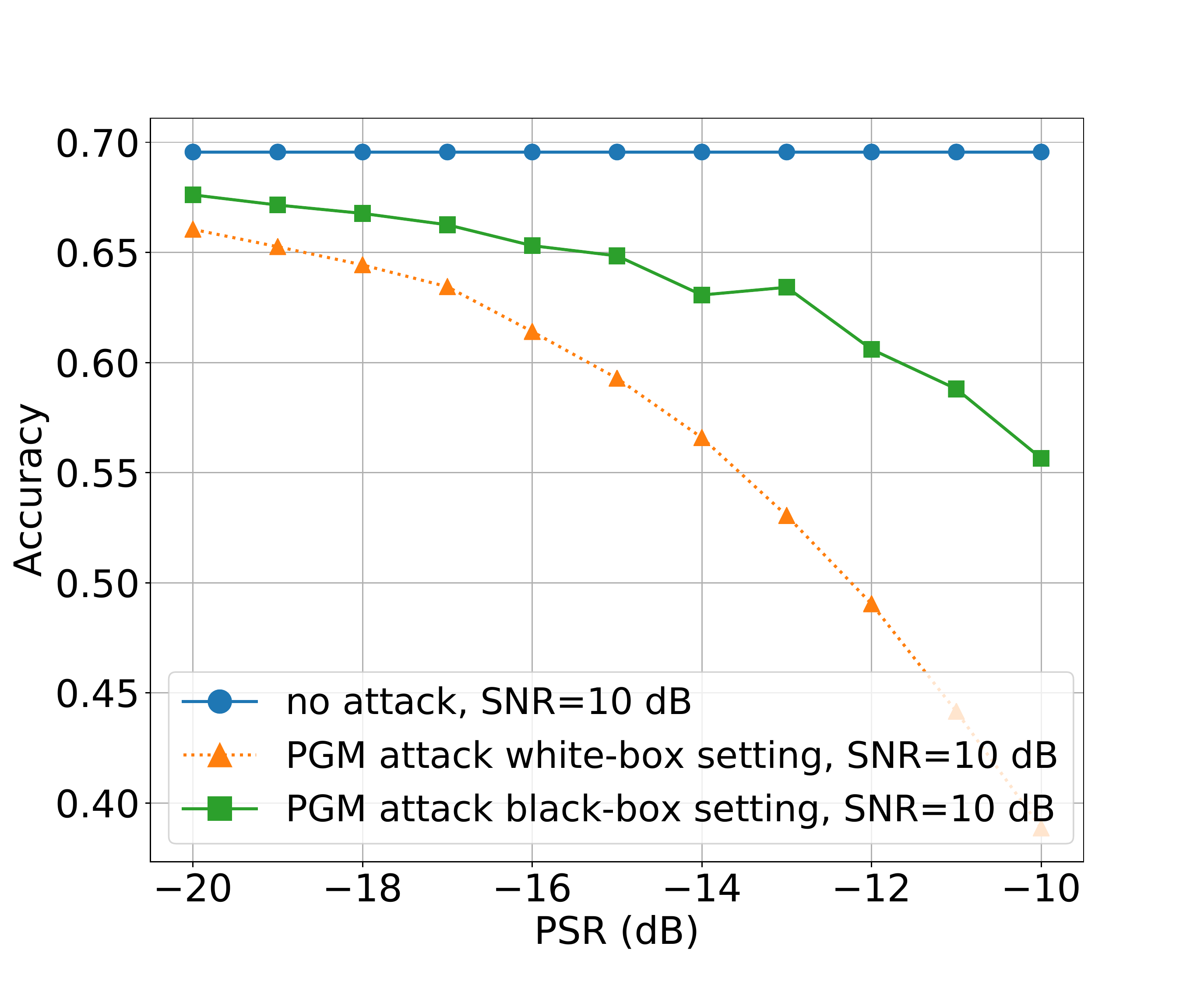}
    \caption{\footnotesize Modulation recognition system}
    \label{fig:attack-modulation-bb}
  \end{subfigure}
  \begin{subfigure}{0.33\textwidth}
    \centering
    \includegraphics[width=\linewidth, trim={1cm, 0cm, 1cm, 1cm}]{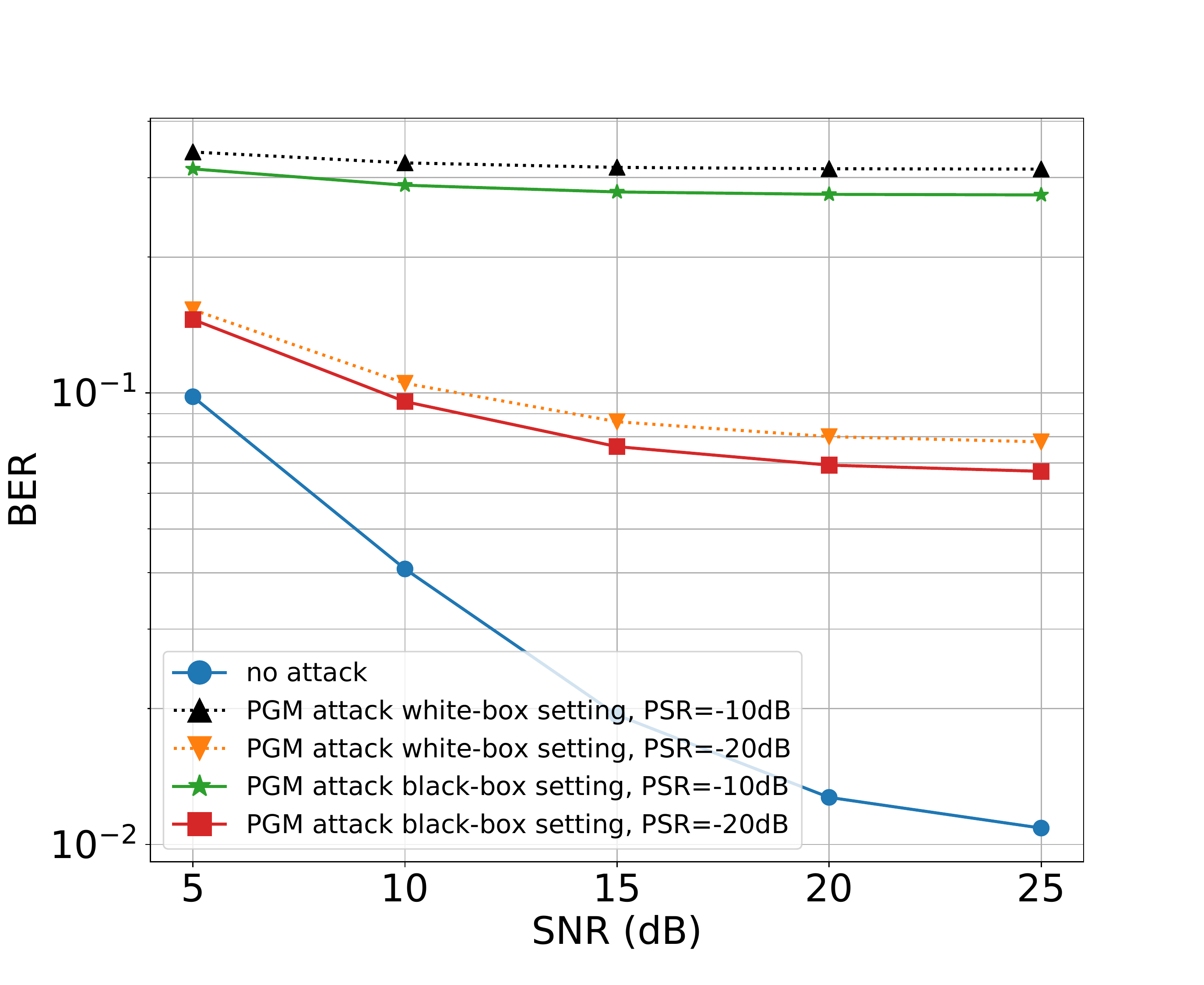}
    \caption{\footnotesize OFDM system}
    \label{fig:attack-ofdm-bb}
  \end{subfigure}
  \caption{Performance of our attack with white-box and black-box scenarios for the three target wireless systems.}
  \label{fig:bb}
\end{figure*}

Using this approach, for each target application, we use a substitute model to design our PGM. We then use the learned PGM to generate perturbations and use them on the original model of each target underlying DNN model.
Note that this approach is general such that the attacker can use any other DNN-based wireless model to generate perturbations and attack the original underlying DNN models.
Table~\ref{tab:params-bb} shows the structure and parameters of the substitute models.
Figure~\ref{fig:bb} compares the performance of our PGM attack in white-box and black-box scenarios for three target wireless applications.
Although in the black-box setting, our PGM attack performs slightly worse than the white-box, we can see that the attack is still effective and degrades the performance of the three underlying DNN models.

\begin{table*}[!t]
\centering
\caption{Details of the substitute models for each wireless application}
\begin{tabular}{|c|cc|c|c|}
  \cline{2-5}
  \multicolumn{1}{c|}{}& \textbf{Autoencoder Encoder} & \textbf{Autoencoder Decoder} & \textbf{Modulation Recognition} & \textbf{OFDM Channel Estimation} \\
  \hline
  input size & 16 & $1\times 2 \times 7$ & 256 & $1\times 4 \times 64$\\
  \hline
  hidden layers sizes & 16, 272 & 112, 32 & 1024, 1024, 512, 128 & 3456, 500, 250, 120\\
  \hline
  hidden layers activations & eLU, \_ & ReLU, \_ & ReLU, ReLU, ReLU, ReLU, Softmax & ReLU, ReLU, ReLU, Sigmoid \\
  \hline
  kernels & 16 & 16, 8 & \_ & 32, 64\\
  \hline
  kernel size & 6 & $(2,3)$, $(2,3)$ & \_ & $(2, 8)$, $(2, 8)$\\
  \hline
  stride & 1 & $(1,1)$, $(1,1)$ & \_ & $(2, 1)$, $(1,1)$\\
  \hline
  padding & 3 & $(1,1)$, $(1,0)$ & \_ & $(0, 1)$, $(0, 1)$\\
  \hline
  convolutional activations & ReLU & ReLU, ReLU & \_ & ReLU, ReLU\\
  \hline
  output size & $2\times 7$ & 16 & 11 & 16\\
  \hline
\end{tabular}
\label{tab:params-bb}
\end{table*}

\section{Countermeasures and Robustness Analysis}\label{defense}
In this section, we propose two countermeasures as defense mechanisms against adversarial attacks in wireless communication systems.  We apply the defenses on both the single vector UAP attack and the PGM attack and evaluate the performance of these attacks.
The performance of the defense mechanisms depends on what the adversary knows about the attack algorithms. In the following, we make assumptions regarding the knowledge of the adversary about both the attacks.

\paragraphb{Single Vector UAP Attack: }
We mentioned before that a single UAP vector can be identified using pilot signals. A defender can transmit known pilot signals and receive the perturbed signal. Since the transmitted signal is known, the defender can subtract it from the received signal which produces an estimate of the perturbation. Therefore, in the single vector UAP attack, we assume that the defender has an estimate of the perturbation.

\paragraphb{Perturbation Generator Model: }
When the attacker uses a PGM, the defender cannot identify the perturbations since the PGM generates a different perturbation vector for each input sample. Hence, in this case, the defender can only obtain knowledge about the PGM or an estimate of the generated noises. Based on this knowledge, we assume two scenarios for the defender:
\begin{itemize}
  \item \textit{Ad-hoc defender: } In this scenario, the defender is not aware of the PGM, and similar to the defense for the single UAP vector attack, the defender uses pilot signals to estimate the generated perturbations, e.g., the defender transmits pilot signals and subtracts them from the received signals, then she takes an average of them to obtain an estimate of the perturbations generated by the PGM.
  \item \textit{Structure-aware defender: } In this scenario, we assume that the defender is aware of the structure of the PGM but not its parameters. Hence, the defender needs to train the PGM on its training data and obtain the learned parameters. We also assume that the defender has the same training dataset as the adversary.
  \item \textit{Perfect-aware defender: } In this scenario, we assume that the adversary is aware of both the structure of the adversary's PGM and its learned parameters also. This is an unlikely assumption in that the defender cannot obtain the learned parameters of the PGM using pilot signals or any other techniques.
\end{itemize}
Note that, in the above mentioned scenarios, we assume that the defender is aware of the power constraint ($p$) of the adversary.

\subsection{The Adversarial Training Defense}
Many defenses have been designed for adversarial examples in image classification applications, particularly, adversarial training, gradient masking, and region-based classification.
In adversarial training~\cite{tramer2017ensemble, madry2017towards, kurakin2016adversarial}, in each iteration of training, this method generates a set of adversarial examples and uses them in the training phase by expanding the training dataset. In our work, we use adversarial training where the defender uses UAPs crafted by our attack to make the target DNN-based wireless model robust against the attacks. The defender trains the DNN-based model for one epoch and then generates blind adversarial perturbations from all possible settings using Algorithm~\ref{alg:main}. Then, she extends the training dataset by including all of the adversarial samples generated by the adversary and trains the DNN-based model on the augmented train dataset.
Algorithm~\ref{alg:def} sketches our defense algorithm.

\begin{algorithm}[t!]
    \caption{Adversarial Training against adversarial attacks in wireless communication systems}
    \label{alg:def}
   \begin{algorithmic}
     \STATE Randomly initialize underlying DNN-based network $N$
     \STATE $\mathcal{D}^{tr} \gets$ training data
     \STATE $\mathcal{L}_{f} \gets $ DNN-based loss function
     \STATE $\mathcal{M} \gets$ domain remapping function
     \STATE $\mathcal{R} \gets$ domain regularizations function
     \STATE $T \gets $ epochs
     \STATE $Z \gets $  [] \MCOMMENT{List of adversarial perturbations}
     \FOR{epoch $t \in \{ 1 \cdots T \}$  }
       \STATE Train the model $N$ for one epoch on training dataset $\mathcal{D}^{tr}$
       \STATE $Z \gets$ generate crafted adversarial samples using generated perturbations by Algorithm~\ref{alg:main} (or the single vector UAP)
       \ENDFOR
       \STATE $\mathcal{D}^{tr}$.extend($\mathcal{D}^{tr} + Z$)
       \RETURN $N$
    \end{algorithmic}
\end{algorithm}

In the case of a single vector UAP attack, the defender uses that single obtained perturbation to generate adversarial samples and train the target model. For the perturbation generator attack, we assume the defender is structure-aware and uses her trained PGM to generate adversarial samples and train the target model.

\subsection{The Perturbation Subtracting Defense}
This is a defense specialized to our domain. In this kind of defense, at the receiver side, the defender performs operations on the perturbed received signal based on her knowledge of the adversary to remove the effect of the perturbation and reconstruct the originally transmitted signal
For the single vector UAP attack, since we assume that the defender has identified an estimate of the perturbation, she can easily subtract her estimate of perturbation from the received signal and obtain the originally transmitted signal.
If the adversary uses a PGM, as mentioned above, we consider three scenarios based on the knowledge of the defender: ad-hoc defender, perfect-aware, and structure-aware defender. In all of the scenarios, once the defender receives the received perturbed signal, she generates a perturbation using her PGM and subtracts it from the received signal.

\begin{figure*}
  \begin{subfigure}{0.33\textwidth}
    \centering
    \includegraphics[width=\linewidth, trim={1cm, 0cm, 1cm, 1cm}]{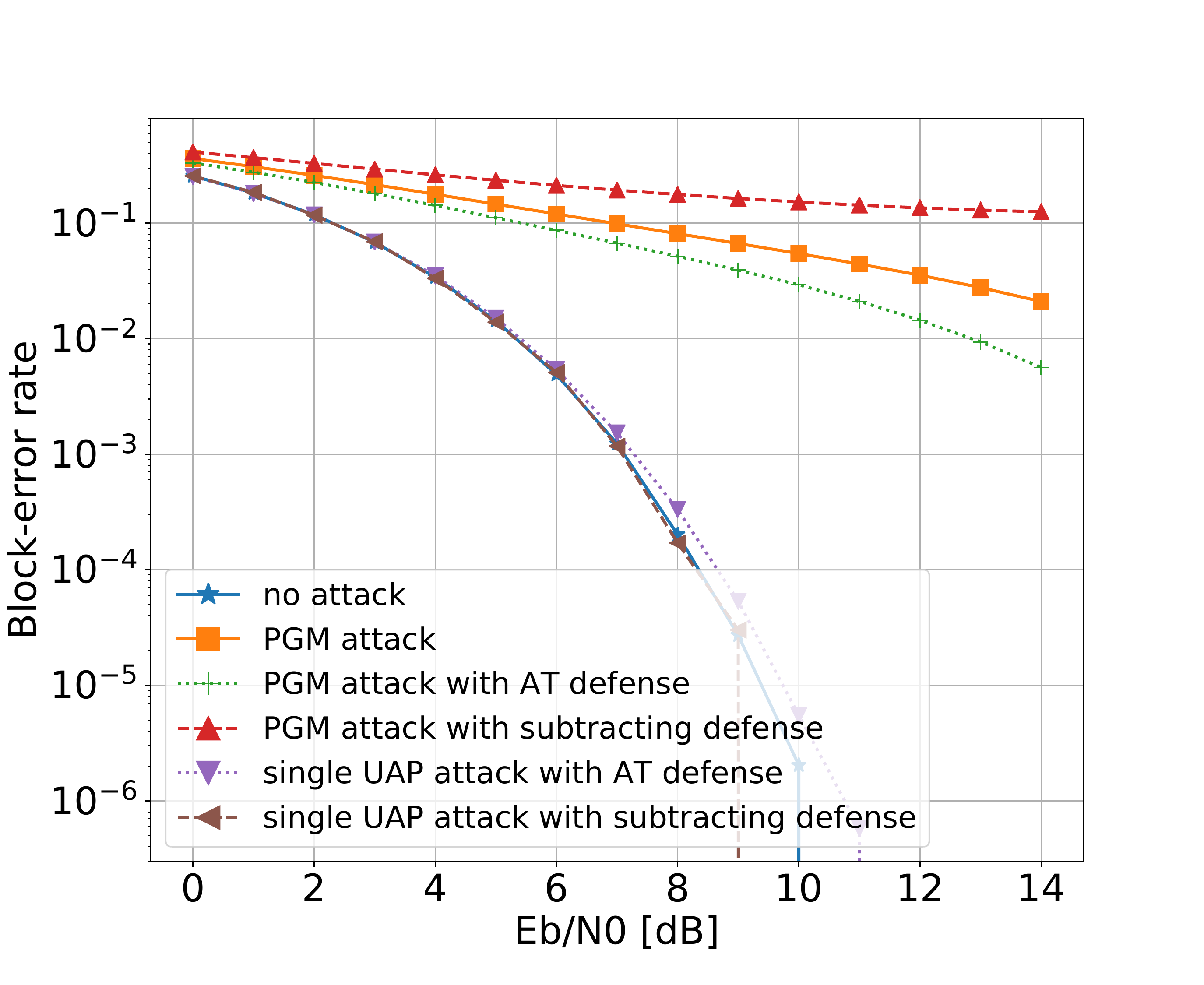}
    \caption{\scriptsize Autoencoder communication system, PSR=$-6dB$}
    \label{fig:sub-autoencoder}
  \end{subfigure}
  \begin{subfigure}{0.33\textwidth}
    \centering
    \includegraphics[width=\linewidth, trim={1cm, 0cm, 1cm, 1cm}]{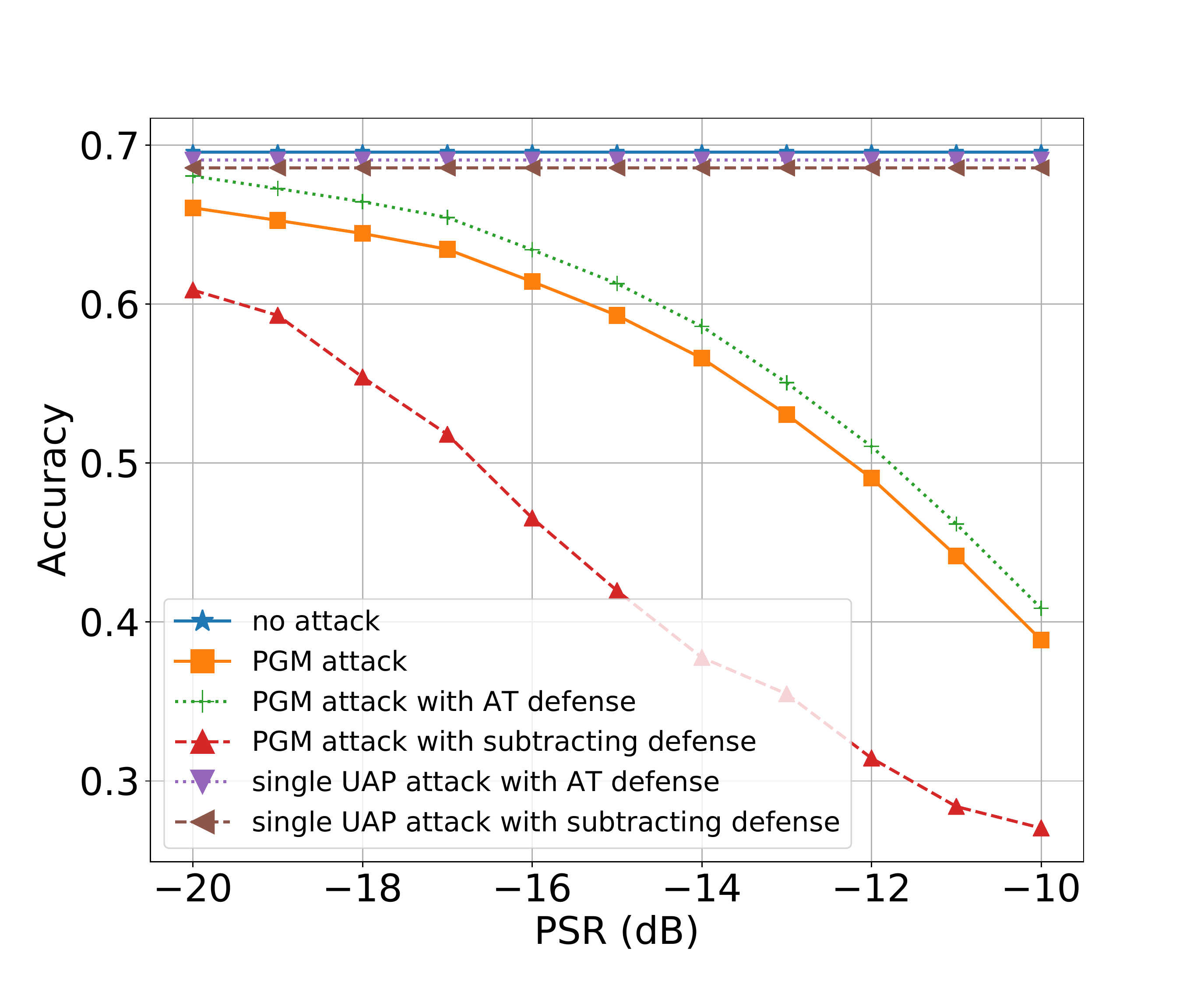}
    \caption{\footnotesize Modulation recognition system, SNR=$10dB$}
    \label{fig:sub-modulation}
  \end{subfigure}
  \begin{subfigure}{0.33\textwidth}
    \centering
    \includegraphics[width=\linewidth, trim={1cm, 0cm, 1cm, 1cm}]{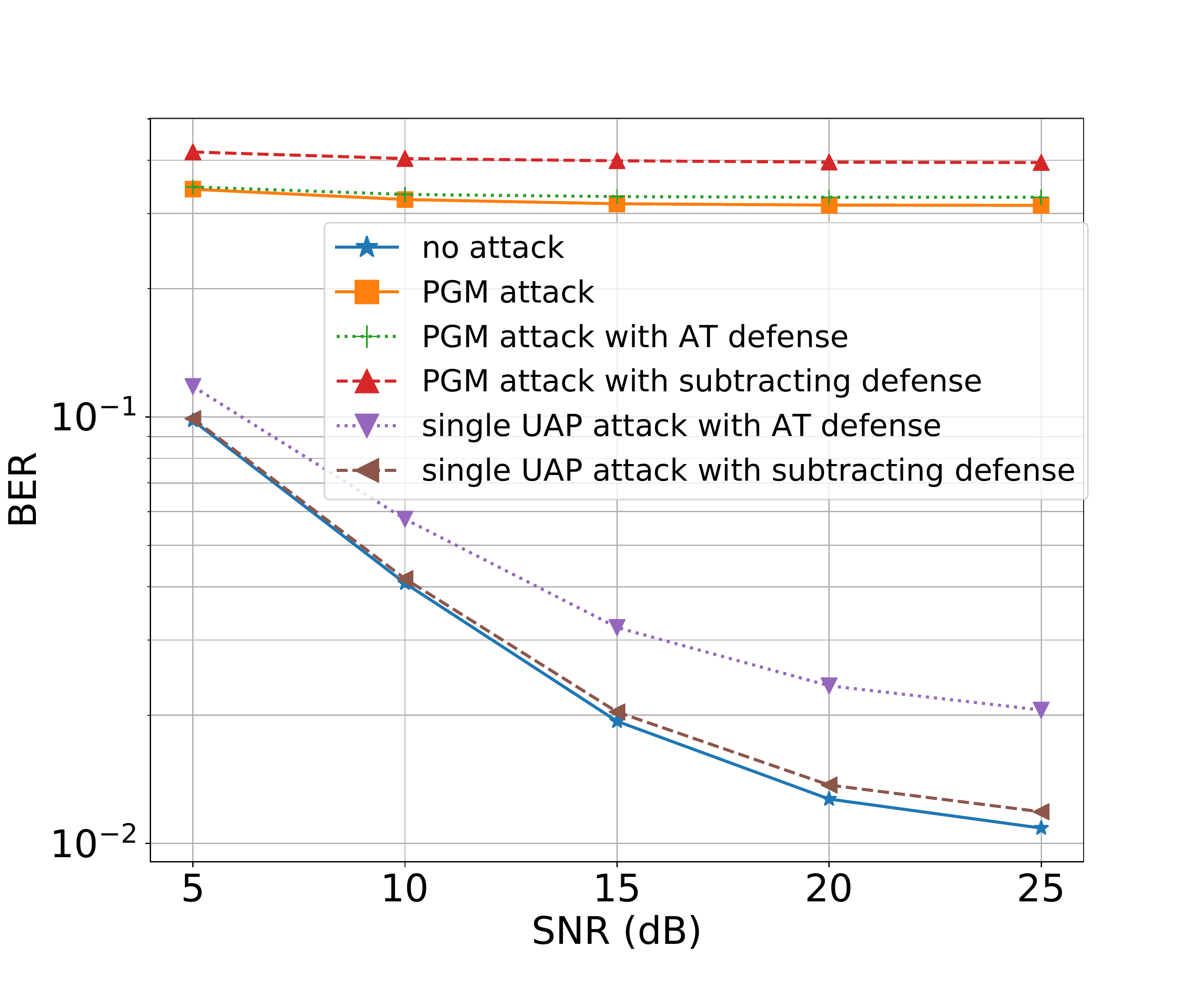}
    \caption{\footnotesize OFDM system, PSR=$-10dB$}
    \label{fig:sub-ofdm}
  \end{subfigure}
  \caption{Performance of the single vector UAP attack and PGM attack against adversarial training and perturbation subtracting defenses.}
  \label{fig:defense}
\end{figure*}

\subsection{Results}
\begin{figure}[t]
  \centering
  \includegraphics[width = 0.8\linewidth]{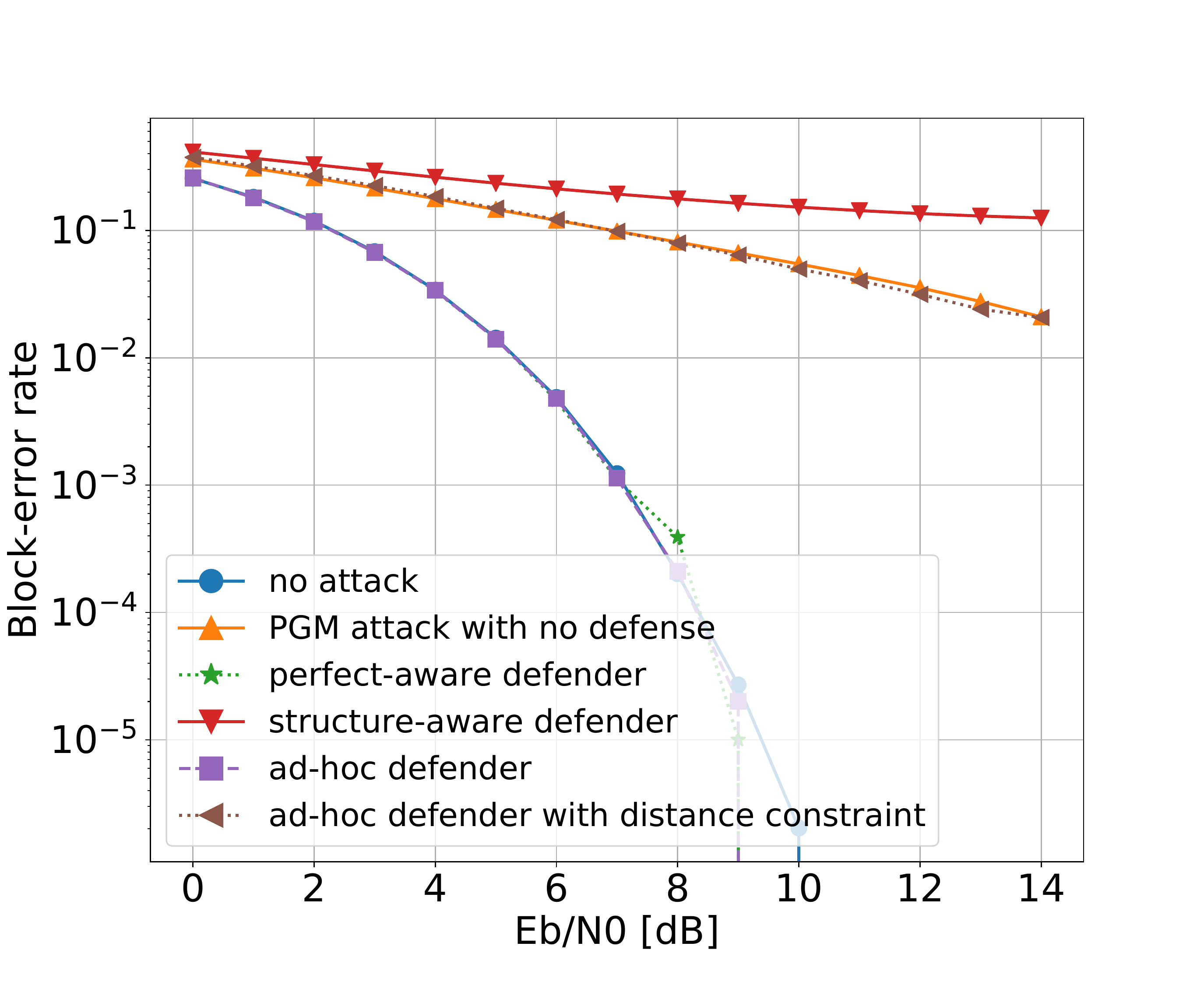}
  \caption{Performance of the autoencoder communication system against the PGM attack with the presence of a perturbation subtracting defense for the structure-aware, perfect aware, and ad-hoc aware defenders.}
  \label{fig:defense-perfect}
\end{figure}

First, we evaluate the performance of our attack against the perturbation subtracting defense in the three above mentioned scenarios. Figure~\ref{fig:defense-perfect} shows the BLER of the autoencoder communication system against the PGM adversarial attack with the presence of the ad-hoc, structure-aware, and perfect-aware defenders. We assume that the power constraint of the attack is $-6dB$.
In the experiments for the ad-hoc defender, we use 10000 pilot signals and take the average of the obtained noise to estimate the perturbation.
 We mentioned that awareness of the learned parameters of the adversarial model is unlikely; however, a perfect-aware adversary can completely degrade the performance of the adversarial attack. Although our adversarial attack provides the attacker a large set of adversarial perturbations, using the same PGM removes the effect of the adversarial attack.
However, a structure-aware defender who trains her PGM not only increases the BLER of the attack but improves the performance of the attack as well.

Furthermore, we see that the ad-hoc defender also degrades the performance of the adversarial attack using pilot signals, which shows that the generated perturbations are too similar to each other such that the defender cancels them by just subtracting a simple estimate of the perturbations. This happens particularly when the adversary uses the same PGM as the one he uses while the defender estimates the generated noises.
To prevent this, the attacker enforces a distance constraint as discussed in Section~\ref{model} to maximize the distance between the generated noise. Figure~\ref{fig:defense-perfect} shows that by enforcing the distance constraint in the training process of the PGM, the attacker removes the effect of the ad-hoc defender. In such a case, the average estimate of the ad-hoc defender is not a precise estimate of the perturbations.
The comparison is similar for other applications and we did not add the results due to space constraints.

We also compare the performance of our attack and the single vector UAP attack against different defense algorithms.
Figure~\ref{fig:defense} shows the performance of our adversarial attack and the single vector UAP attack against the mentioned defense methods. Note that in both defense methods, we only consider the structure-aware defender where the defender only knows the structure of the PGM and obtains her learned parameters. Furthermore, as mentioned above, in the single vector UAP attack scenario, we assume that the defender can use pilot signals to estimate the single perturbation. In our experiments, we obtain this estimate by sending 10000 pilot signals and averaging their resulted perturbation. Hence, in the adversarial training defense, we use only the estimated perturbation to generate adversarial samples and train the target DNN-based model.

Figures~\ref{fig:sub-autoencoder}, \ref{fig:sub-modulation}, \ref{fig:sub-ofdm} show that the PGM attack is more robust against the defense mechanisms.
We see that subtracting the estimated noise from the received signal at the receiver side destroys the effect of the attack completely, while even a structure-aware adversary not only reduces the effect of the PGM attack but worsens the performance of the DNN-based model also. Hence, the perturbation subtracting defense method is more effective on the single vector UAP attack in comparison to the PGM attack.

Based on these figures, we can say that the AT defense is less effective than perturbation subtracting, however, our attack is still robust against the AT defense compares to the single vector UAP attack.
The reason is that the underlying DNN-based model can learn the adversarial samples generated just by a single perturbation. Using a PGM enables the attacker to access an extremely large set of perturbations which makes it infeasible for the defender to learn the DNN-based model based on all of them.
\textbf{Based on our results, we conclude that using a PGM to perform the adversarial attack against wireless communication systems is more robust against various defense mechanisms in comparison to using only a single perturbation vector.}



\section{Conclusions}


In this paper, we propose an adversarial attack using a perturbation generator model against DNN-based wireless communication systems. In the absence of countermeasures deployed by the communicating parties, our attacks perform slightly better than existing adversarial attacks.  More importantly, against communicating parties employing defenses, our technique is more robust and shows significant gains over previous approaches.
We also show that our attack is effective in a black-box scenario where the attacker generate the adversarial perturbations using a substitute DNN wireless model and use the perturbation to attack the original DNN wireless model. 
We use remapping and regularizer functions to enforce an undetectability constraint for the perturbations, which makes them indistinguishable from random Gaussian noise. Furthermore, we use a regularizer function to enforce a distance constraint to degrade the performance of an ad-hoc defender who tries to obtain
an estimation of the perturbations using pilot signals.
Our work shows that even in the presence of defense mechanisms deployed by communication parties, DNN-based wireless systems are highly vulnerable to adversarial attacks.



\bibliographystyle{IEEEtranS}
\bibliography{ref}


\end{document}